# Robust Sign Language Recognition System Using ToF Depth Cameras

Morteza Zahedi and Ali Reza Manashty
Department of Computer Engineering and IT
Shahrood University of Technology
Shahrood, Iran
zahedi@shahroodut.ac.ir and a.r.manashty@gmail.com

*Abstract*—Sign language recognition is a difficult task, yet required for many applications in real-time speed. Using RGB cameras for recognition of sign languages is not very successful in practical situations and accurate 3D imaging requires expensive and complex instruments. With introduction of Time-of-Flight (ToF) depth cameras in recent years, it has become easier to scan the environment for accurate, yet fast depth images of the objects without the need of any extra calibrating object. In this paper, a robust system for sign language recognition using ToF depth cameras is presented for converting the recorded signs to a standard and portable XML sign language named SiGML for easy transferring and converting to real-time 3D virtual characters animations. Feature extraction using moments and classification using nearest neighbor classifier are used to track hand gestures and significant result of 100% is achieved for the proposed approach.

## I. INTRODUCTION

Emerging technologies help scientific researches both in speed and accuracy. Sometimes, previous research efforts on a particular area might become completely or partially out of date with the new technology replacing the previous one. Time-of-Flight (ToF) or Range cameras are type of cameras that are able to create distance images from the environment, ranging from few meters to several kilometers. These devices have been shipped just since the first decade of 21th century when the semiconductor products became fast enough to provide the speed needed for processing the data of such cameras. Recently with current technology, the price of these cameras is reduced about twenty times. This efficiency in the price makes it possible to use such a technology in ordinary life of people when they are handy. The new capabilities of such devices make it possible to track gestures and environments from another point of view that reduces the recognition error of previously used RBG cameras and sophisticated 3D sensing systems.

On the other hand, voice is the primary means of communication between human beings; yet, it is not possible to always rely on voice for communication. This disability might be caused from different situations, e.g. people with auditory defects that also lost their ability to speak or just prefer to do so, military situations that voice communication may lead to revealing the stealth of the troops and ordinary situations when signs are preferred to voice communication such as in audio recording studios. These reasons, made a lot of people learn the sign language and create a standard for it and they even created a language for it. American Sign Language (ASL) is one example of a commonly used sign language.

Many projects and efforts are aimed for removing the gap between the signers and regular speakers [1] using regular cameras. These efforts sometimes involve creating new databases or modifying and discussing the older ones hoping for a better recognition rate [2, 3]. The use of depth in recognition is mostly limited to stereo imaging and the results are not satisfactory for general public usage [4].

To transfer a sign sentence via mediums to another place, the easiest way is to transfer the video sequence of the recorded sign language over the transmission medium to the destination so that a person on the other side of this communication can interpret the meaning of the sign language. Although this seems very simple, the video transmission requires high bandwidth to be transferred. Due to the explained problem, a standard sign language that can be used as a middle coding for transferring such data is devised and is named SiGML (Signing Gesture Markup Language) [5]. This language is still in use in recent researches [6]. SiGML is an XML language that is defined based on the single movements of the body, from eyes to fingertips. This language that is also compatible with the Hamburg Sign Language Notation System (or HamNoSys, a phonetic transcription system for sign languages) is used in most research studies.

In this paper, we present the use of ToF depth cameras to track different hand movements and after recognition of the single movements, convert them to the appropriate SiGML notation so that it can be easily transferred as a text and then the same recognized action can be shown using 3D virtual characters. Different steps toward such system are organized in this paper as sections so after previous works on the next section, section 3 is about data acquisition and database, section 4 describes the preprocessing steps, section 5 and 6 discuss the feature extraction and classification schemes using nearest neighbor (NN) classifier, respectively, and then the paper is concluded with the results and future works.

## II. PREVIOUS WORKS

The need for storing and transferring the sign data makes different sign notation systems to evolve. Stokoe system, HamNoSys and SignWriting are some of those [7]. From the above, The Hamburg Sign Language Notation System, HamNoSys [8], is a "phonetic" transcription system, which is a more general form of Stokoe system. It includes more than 200 symbols which are used to transcribe signs in four levels consisting of hand shape, hand configuration, location and movement. The symbols are based on iconic symbols to be easily recognizable. Facial expressions can be represented as well, but their development is not finished yet. HamNoSys is still being improved and extended as the need arises and hence is the most suitable for converting real sign gestures to the equivalent notation.

The Signing Gesture Markup Language, SiGML [5] is based on HamNoSys symbols and represents signs in form of Extensible Markup Language (XML) – a simple but flexible format for transferring structured and semi-structured data. There are some tools available to convert signs written in SiGML format to virtual character animations, both as applications in regular operating systems and as applets in web browsers. These tools make SiGML a preferable language for converting the recognized signs to it.

For robust gesture recognition, in most of the efforts, the equipment used in order to capture the videos is usually complex and expensive. For example, authors in [9] use two cameras for recording or authors in [10] use dense ranging images captured by expensive 3D sensors. Others usually use some equipment other than cameras, such as special sensor gloves, like [11] and [12]. The use of regular and efficient ToF cameras which can be easily purchased by individuals is not a habit in sign language recognition. Specially the efforts to use ToF cameras that can obtain the data remotely, accurately and fast for converting the signs to SiGML language (so that it can be easily transferred to be visual again using virtual animated characters) has not been tested before. The latter effort helps people to communicate over networks with lower bandwidth or use short text messages systems on cellular phones to transfer, convert and watch the animated sign language representation of the text message easily.

In this paper, a novel approach is proposed for recognizing the signs that are captured using a ToF depth camera and converting the recognized signs to SiGML language. The signs in this language can be easily replayed using SiGML service player that uses a 3D virtual character animation to replay the recorded signs.

## III. DATA ACQUISITION

To acquire data for recognition, a ToF camera similar to Microsoft Kinect for Xbox 360 is used. The output of these cameras is depth sequence of images (video) in gray-scale at about 30 frames per second. The images show the environment in such a way that the nearest objects have more gray-scale intensity and are close to white and farther objects have lower gray-scale intensity and are close to black.

### A. Gestures to Capture

The SiGML language has many elements that can be used to define a movement. This language also supports HamNoSys elements to be used. For a robust start, four different movements have been chosen that are to be recognized. These movements are listed below on Table 1. As it can be observed, each movement has a symbol in HamNoSys font and system and has an equivalent in SiGML language too. The point in these four different movements is that they are not started in a particular area of the screen and are not even finished in a predefined area. They can be started anywhere in the screen and be finished anywhere else, so all different possible starting and ending point for these gestures must be included in the database.

TABLE 1.  THE FOUR DIFFERENT HAND MOVEMENTS IN THE DATABASE

| Movement | Equivalent Symbols | | |
|---|---|---|---|
| | *SiGML*[a] | *HamNoSys Unicode* | *HamNoSys Symbol* |
| Hand to Right | hammover | E082 | → |
| Hand to Left | hammovel | E086 | ← |
| Hand to Up | hammoveu | E080 | ↑ |
| Hand to Down | hammoved | E084 | ↓ |

a. These elements are special HamNoSys equivalent elements.

### B. Database

For creating a suitable and fault-tolerant database, the video of nine different sets of all four previously mentioned movements is captured. This results in 36 single movements. These videos are captured using a ToF camera in 640x480 resolution and maximum of 30 frames per second. The database includes every action in one consecutive video but a single blank action is recorded between every set of actions. Single actions are separated using the two steps described in the preprocessing section.

## IV. PREPROCESSING

The raw video data that is captured using the ToF camera needs some processing so that it can be used for classification. As it has been mentioned earlier, the range image frames are gray-scale with the whiter pixels meaning they are nearer to the camera. Most of the gesture recognition problems were that the recognition of the skin and the hand itself needed some noticeable amount of time and complexity. The unavoidable errors caused by lighting and noise made it difficult to track the hand itself. When the output image of the ToF camera is used, without any effect of light, the results have the same characteristics in any room lighting condition (because of the nature of the ToF cameras, sunlight defects the received light of such devices).

*A. Intensity Filtering*

The first preprocessing step applied to the video frames is to threshold the pixels to a specific range, resulting in all images (video frames) with the real-world distance filter applied to. In this research, the pixels with intensity of less than 112 and more than or equal to 128 have been filtered. This means that only a domain of 16 pixel values are selected for the images and the others turned to black. This filtering makes it possible to have only the hand in the screen and have every other useless details removed very fast and accurately. In an automatic manner, these numbers can change automatically depending on the distance of the person but in this research, the distance of the person is changed so that when he adjusts his hand to be visible in the filtered visible range of the camera, the person does not move during the whole database creation period. While having little varying z coordinate, each set of actions has completely different x and y coordinates.

Fig. 1 shows the starting and ending frame of a regular gesture video taken from ToF camera after filtering. The images are gray-scale and for convenient, some color spectrum has been used to clarify it. Fig. 2 and Fig. 3 show images of an action that is not performed near to the center of the screen and a defected captured action, respectively. Both the actions shown in Fig. 2 and Fig. 3 are included in the database to show the robustness of the recognition.

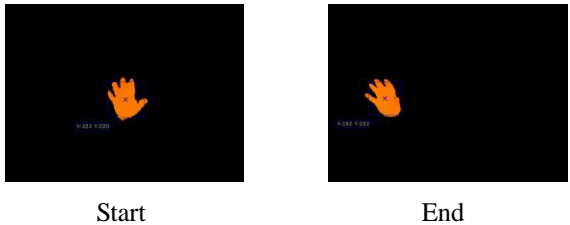

Start             End

Figure 1. The starting and ending frame of a right hand movement after filtering. The center of gravity of the hand is also calculated.

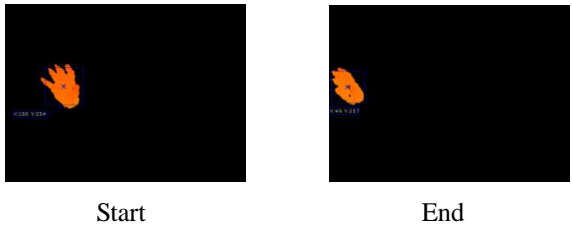

Start             End

Figure 2. The starting and ending frame of a right hand movement that is not performed near to the center of the screen; after filtering. The center of gravity of the hand is also calculated.

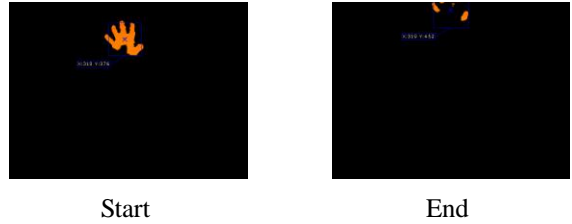

Start             End

Figure 3. The starting and ending frame of a defected captured image showing hand movement to the top; after filtering. The center of gravity of the hand is also calculated despite most of the hand is not visible.

*B. Action Separation*

The second preprocessing step that can also be automatically done later, is to separate single movements in continuous set of actions. This is done by (1) separating each single action frames from all the video frames and (2) selecting the starting and ending image of each sequence. The second step described above is performed due to nature of the actions in the database that is selected for classifying. If a more complex action is going to be recognized, many inner frames other than the starting and ending frame will be needed and some other algorithms and tools such as hidden Markov model can be used.

V. FEATURE EXTRACTION

After the preprocessing step, two images per action are remained, showing the starting and ending frame of that action. Each of these images includes the hand of the signer fully or partially. To classify these patterns correctly, moments are used to get the features from the images:

*A. Moments*

The features that are selected for classification are based on moments [13]. The two dimensional (p+q)th order moments of the grey-value image with the pixel intensities X(i, j) are defined as following:

$$m_{pq} = \sum_{i}^{I}\sum_{j}^{J} i^p j^q X(i, j) \qquad (1)$$

If X(i, j) is the piecewise continuous and it only has non-zero values in the finite part of the two dimensional plane, then the moments of all orders exist and the sequence {$m_{pq}$} is uniquely determined by X(i, j) and vice versa. The small order moments of the X(i, j) describe the shape of the region. For example $m_{00}$ is equal to the area size, and $m_{01}$ and $m_{10}$ give the x and y coordinates of the center of the gravity [7].

*B. Feature Selection*

As the criteria for using moments are met in our database images, two features are chosen from the moments of the images: $m_{01}$ and $m_{10}$ which are called $x_{CoG}$ and $y_{CoG}$ as the x and y coordinates of the center of gravity of the hand image. For every action, there are two images, one for the starting image of the action and the other for the ending image of the same action. So the total number of features for any action is equal to four, two for the center of gravity of the starting image and two for the center of gravity of the ending image.

Nearest neighbor classifier is used for the classification of the actions. First, it must be assured the input features are discriminant enough to result in a good classification. To ensure that, the scatter diagram of the center of gravity points of both the starting and ending images has been plotted separately. As it is visible in the Fig. 4 and Fig. 5, because the starting and ending point of each set of action is not at the center of the screen, each movement could start and end anywhere in the screen so that it may be misinterpreted for another hand movement. Because these features are indiscriminant, instead of using the center of gravities as features, the movement vector of the center of gravity of each action, resulted from the following equation, is used:

$$(x, y)_{\text{CoG-Movement Vector}} = (x_{\text{CoG-End}} - x_{\text{CoG-Start}}, y_{\text{CoG-End}} - y_{\text{CoG-Start}}) \quad (2)$$

In (2), as the resulting vector is two dimensional, the total number of features is reduced by two. The scatter diagram of the center of gravity movement vector calculated above is illustrated in Fig. 6. As it can be easily observed, each action is now scattered in a particular part of the space and is now easily classifiable by nearest neighbor. Each hand movement vector in Fig. 6 is cumulated with its similar hand movement vectors, all in their particular part of the space.

## VI. CLASSIFICATION

The feature vector created from the previous step consists of two values for each action. As movement vector of the center of gravity is used, the extracted features are scattered in a discriminant way so that they can be easily classified using the nearest neighbor algorithm. The nearest neighbor classifier used for the classification utilizes Euclidian distance measure to determine the distance between each sample.

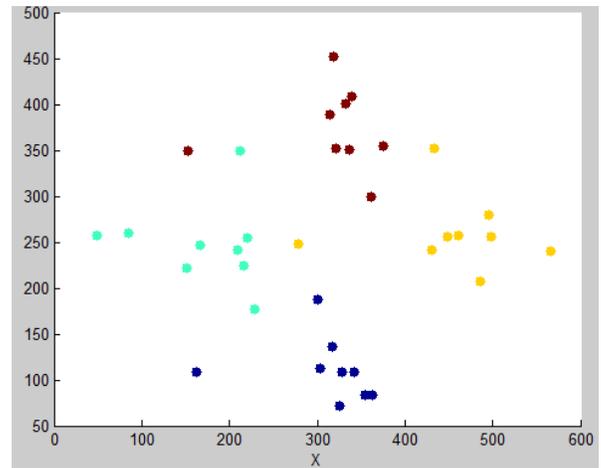

Figure 5. The scatter diagram of the center of gravity points of the ending image of different actions. The colors of the points that correspond to each action are dark Red, Blue, Yellow and Cyan, for movement to Top, Bottom, Right and Left, respectively.

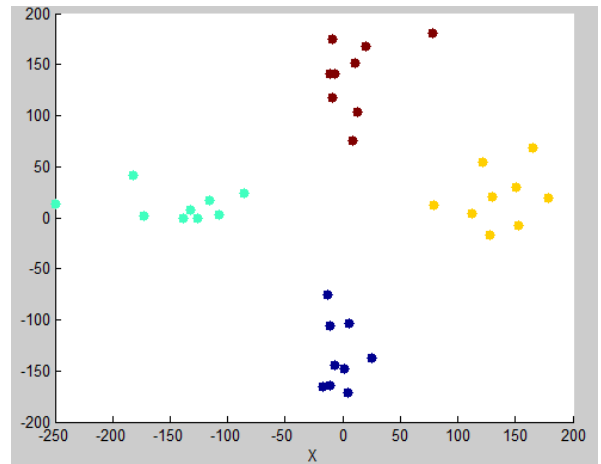

Figure 6. The scatter diagram of the movement vectors of the center of gravity points of the actions. The colors of the points that correspond to each action are dark Red, Blue, Yellow and Cyan, for movement to Top, Bottom, Right and Left, respectively. The points are scattered in their groups and are discriminant enough for classification using nearest neighbor algorithm.

### A. Training and Testing

From nine sets of four different hand movements, 5 sets are used for training and four sets for testing. This results in 20 reference samples and 16 test samples. Using nearest neighbor, each of 16 test samples are compared with 20 train samples and the train sample with the least distance will determines the test sample's class. Each test sample must be grouped to one of the following four classes: Movement to Right, Movement to Left, Movement to Top and Movement to Bottom.

### B. Classification Results

The permutation of five training sets that can be chosen from nine total samples is equal to 126. The train sets have been changed 126 times so that all possible combination of train and test data can be verified for consistency. In all 126 classifications of hand gesture movements using nearest neighbor classification, 100% classification result is achieved.

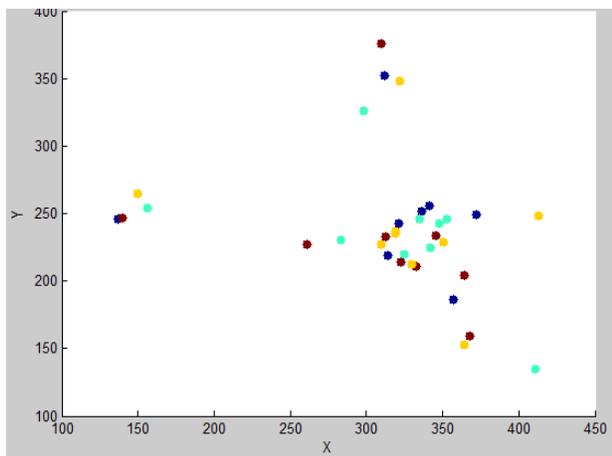

Figure 4. The scatter diagram of the center of gravity points of the starting image of different actions. The colors of the points that correspond to each action are dark Red, Blue, Yellow and Cyan, for movement to Top, Bottom, Right and Left, respectively.

## VII. GESTURE TO SiGML TRANSLATION

After successfully classifying the input hand gestures, according to the Table 1, each successful recognized action can be converted to the corresponding SiGML element. Using a template for a SiGML file, the elements are added one after another after classification. The resulting SiGML file becomes ready for being played in the SiGML service player [14]. The SiGML service player can play SiGML files and converts the movements defined in the file to a 3D virtual character animation. An example for resulting animation of the recognized real human gestures that is being played in the SiGML service player is shown in Fig. 7.

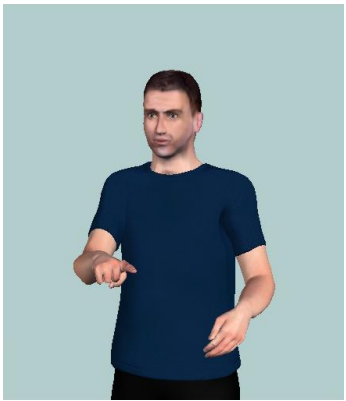

Figure 7. An example of resulting animation of the recognized real human gestures that is being played in the SiGML service player

## VIII. CONCLUSION AND FUTURE WORKS

In this paper, the effective role of ToF depth cameras in making the sign language recognition a faster and more reliable task is presented. The importance of sign language is clear both as a means of communication for disabled people and also as a second language for situations in which communication using voice is not possible.

The transmission of performed sign languages is important since transferring the input video might be too costly and space consuming. On the other hand, relying on individuals for interpreting the signs is not always possible and efficient. So some sign writing languages are devised to transfer the recognized signs, e.g. SiGML and HamNoSys notation systems. The aim of this paper is to present a way to recognize the input video of performed signs and to convert them in the easiest and most accurate way to SiGML language for easy transferring through the media. The robustness of the features selected for the classification is approved with the result of 100% for the selected gestures.

In near future, the database must be expanded to cover the most useful and common symbols and actions of the sign language to propose a system that is capable of recognizing more complex and handy signs. Different aspects of signing must be considered in such tasks [15]. There is also the need for a model that can help the translation of more advanced sign actions to be an easier job. The automation of distance calibrating and action separation is also in high priority.

ACKNOWLEDGMENT

We sincerely thank Zahra Forootan Jahromi for helping us in creating the database and testing the ToF camera in the first place.

AUTHORS PROFILE

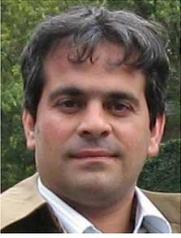

Morteza Zahedi received the B.Sc. degree in computer engineering (hardware) from Amirkabir University of Technology, Iran, in 1996, the M.Sc. in machine intelligence and robotics from University of Tehran, Iran, in 1998 and the Ph.D. degree in man-machine interaction from RWTH-Aachen University, Germany, in 2007, respectively. He is currently an assistant professor in Department of Computer Engineering and IT at Shahrood University of Technology, Shahrood, Iran. He is the Head of Computer Engineering and IT Department. His research interests include pattern recognition, sign language recognition, image processing and machine vision.

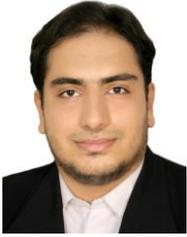

Ali Reza Manashty is a M.Sc. student of artificial intelligent at Shahrood University of Technology, Shahrood, Iran. He received his B.Sc. degree in software engineering from Razi University, Kermanshah, Iran, in 2010. He has been researching on mobile application design and smart environments especially smart digital houses since 2009. His publications include 6 papers in international journals and conferences and one national conference paper. He has earned several national and international awards regarding mobile applications developed by him or under his supervision and registered 4 national patents. He is also a member of The Elite National Foundation of Iran. His research interests include 3D image processing, pattern recognition, smart enviroments and mobile agents.